\documentclass{aastex631}
\usepackage{amsmath}

\def\mybf{}

\def\msun{$M_{\odot}$}

\def\sqig{$\sim$}

\def\degrees{$^{\circ}$}
\def\ctscm2s{cts\,cm$^{-2}$\,s$^{-1}$}
\def\ctss{cts\,s$^{-1}$}

\def\ergscm2s{ergs\,cm$^{-2}$\,s$^{-1}$}
\def\ergss{ergs\,s$^{-1}$}

\def\Swift{{Swift}}

\newcommand {\PPdot} {$P/\dot{P}$}

\def\src{XTE\,J1829$-$098}
\def\a0538{A 0538-66}

\begin{document}
\accepted{October 3, 2024}
\journalinfo{Astrophysical Journal}

\title{Sharp Periodic Flares and Long-Term Variability in the High-Mass X-ray Binary XTE J1829$-$098 from RXTE PCA, Swift BAT and MAXI  Observations
}

\author[0000-0002-3396-651X]{Robin H.~D. Corbet}
\affiliation{University of Maryland, Baltimore County, 1000 Hilltop Cir, Baltimore, MD 21250, USA}
\affiliation{X-ray Astrophysics Laboratory, Code 662 NASA Goddard Space Flight Center, Greenbelt Rd., MD 20771, USA}
\affiliation{CRESST II}
\affiliation{
Maryland Institute College of Art, 1300 W Mt Royal Ave, Baltimore, MD 21217, USA}

\author[0000-0002-1118-8470]{Ralf Ballhausen} 
\affiliation{Department of Astronomy, University of Maryland, College Park, College Park, MD 20742, USA}
\affiliation{X-ray Astrophysics Laboratory, Code 662 NASA Goddard Space Flight Center, Greenbelt Rd., MD 20771, USA}

\author[0000-0002-3718-1293]{Peter A. Becker} 
\affiliation{Department of Physics and Astronomy, George Mason University, Fairfax, VA 22030-4444, USA}

\author[0000-0001-7532-8359]{Joel B. Coley} 
\affiliation{Department of Physics and Astronomy, Howard University, Washington, DC 20059, USA}
\affiliation{CRESST II}

\author{Felix Fuerst} 
\affiliation{European Space Agency (ESA), European Space Astronomy Centre (ESAC), Camino Bajo del Castillo s/n, 28692 Villanueva de
la Cañada, Madrid, Spain}

\author[0000-0001-7115-2819]{Keith C. Gendreau} 
\affiliation{X-ray Astrophysics Laboratory, Code 662 NASA Goddard Space Flight Center, Greenbelt Rd., MD 20771, USA}

\author[0000-0002-6449-106X]{Sebastien Guillot} 
\affiliation{IRAP, CNRS, 9 avenue du Colonel Roche, BP 44346, F-31028 Toulouse Cedex 4, France}

\author[0000-0002-2413-9301]{Nazma Islam} 
\affiliation{University of Maryland, Baltimore County, 1000 Hilltop Cir, Baltimore, MD 21250, USA}
\affiliation{X-ray Astrophysics Laboratory, Code 662 NASA Goddard Space Flight Center, Greenbelt Rd., MD 20771, USA}

\author[0000-0002-6789-2723]{Gaurava Kumar Jaisawal} 
\affiliation{DTU Space, Technical University of Denmark, Elektrovej 327-328, DK-2800 Lyngby, Denmark}

\author[0000-0002-9877-6768]{Peter Jenke} 
\affiliation{NASA Marshall Space Flight Center, Huntsville, AL, USA}
\affiliation{University of Alabama in Huntsville, Huntsville, AL, USA}

\author[0000-0001-9840-2048]{Peter Kretschmar} 
\affiliation{European Space Agency (ESA), European Space Astronomy Centre (ESAC), Camino Bajo del Castillo s/n, 28692 Villanueva de
la Cañada, Madrid, Spain}

\author[0000-0003-3540-2870]{Alexander Lange}
\affiliation{University of Maryland, Baltimore County, 1000 Hilltop Cir, Baltimore, MD 21250, USA}
\affiliation{X-ray Astrophysics Laboratory, Code 662 NASA Goddard Space Flight Center, Greenbelt Rd., MD 20771, USA}
\affiliation{CRESST II}
\affiliation{Department of Physics, The George Washington University, 725 21st Street, NW, Washington, DC 20052, USA}

\author[0000-0002-0380-0041]{Christian Malacaria} 
\affiliation{International Space Science Institute, Hallerstrasse 6, 3012 Bern, Switzerland}
\affiliation{INAF-Osservatorio Astronomico di Roma, Via Frascati 33, I-00078, Monteporzio Catone (RM), Italy}

\author[0000-0002-0940-6563]{Mason Ng}
\affiliation{MIT Kavli Institute for Astrophysics and Space Research, Massachusetts Institute of Technology, Cambridge, MA 02139, USA}

\author[0000-0002-4656-6881]{Katja Pottschmidt}
\affiliation{Astroparticle Physics Laboratory, Code 661 NASA Goddard Space Flight Center, Greenbelt Rd., MD 20771, USA}
\affiliation{CRESST II}

\author[0000-0002-1131-3059]{Pragati Pradhan} 
\affiliation{Department of Physics, Embry Riddle Aeronautical University, Prescott Campus, 3700 Willow Creek Road, Prescott, AZ 86301, USA}

\author[0000-0002-5297-5278]{Paul S. Ray} 
\affiliation{Space Science Division, Naval Research Laboratory, Washington, D.C. 20375, USA}

\author{Richard E. Rothschild} 
\affiliation{Astronomy and Astrophysics Department, University of California San Diego, La Jolla, CA 92093, USA}

\author[0000-0001-6269-2821]{Philipp Thalhammer} 
\affiliation{Dr. Karl Remeis-Observatory and Erlangen Centre for Astroparticle Physics, Universit{\"a}t Erlangen-Nürnberg, Sternwartstr. 7, 96049 Bamberg, Germany}

\author[0000-0001-9788-3345]{Lee J. Townsend}
\affiliation{South African Astronomical Observatory, PO Box 9, Observatory, Cape Town 7935, South Africa}
\affiliation{Southern African Large Telescope, PO Box 9, Observatory, Cape Town 7935, South Africa}
\affiliation{Department of Astronomy, University of Cape Town, Private Bag X3, Rondebosch 7701, South Africa}

\author[0000-0003-2065-5410]{Joern Wilms} 
\affiliation{Dr. Karl Remeis-Observatory and Erlangen Centre for Astroparticle Physics, Universit{\"a}t Erlangen-Nürnberg, Sternwartstr. 7, 96049 Bamberg, Germany}

\author[0000-0002-8585-0084]{Colleen A. Wilson-Hodge} 
\affiliation{ST12 Astrophysics Branch, NASA Marshall Space Flight Center, Huntsville, AL 35812, USA}

\author[0000-0002-4013-5650]{Michael T. Wolff}
\affiliation{Space Science Division, U.S. Naval Research Laboratory, Washington, DC 20375-5352, USA}

\begin{abstract}

\src\ is a transient X-ray pulsar with a period of \sqig7.8 s. It is a candidate Be star system, although the evidence for this is not yet definitive.
We investigated the twenty-year long X-ray light curve using the Rossi X-ray Timing Explorer Proportional Counter Array (PCA), Neil Gehrels Swift Observatory Burst Alert Telescope (BAT), and the Monitor of All-sky X-ray Image (MAXI). We find that all three light curves are clearly modulated on the \sqig244 day orbital period previously reported from PCA monitoring observations, with outbursts confined to a narrow phase range. 
The light curves also show that \src\ was in an inactive state between approximately December 2008 and April 2018 and no strong outbursts occurred.
Such behavior is typical of Be X-ray binary systems, with the absence of outbursts likely related to the dissipation of the Be star’s decretion disk. The mean outburst shapes can be approximated with a triangular profile and, from a joint fit of this to all three light curves, we refine the orbital period to 243.95 $\pm$ 0.04 days. The mean outburst profile does not show any asymmetry and has a total phase duration of 0.140 $\pm$ 0.007. However, the PCA light curve shows that there is considerable cycle-to-cycle variability of the individual outbursts. 
We compare the properties of \src\ with other sources that show short phase-duration outbursts, in particular GS 1843-02 (2S 1845-024) which has a very similar orbital period, but longer pulse period, and whose orbit is known to be highly eccentric.

\end{abstract}


\section{Introduction} \label{sect:intro}

High-mass X-ray binaries (HMXBs) consist of a compact object, a neutron star or a black hole, accreting from an O- or B-star companion, that is the primary star. Mass transfer can occur by either Roche-lobe overflow by the primary, accretion from the stellar wind of a supergiant, or from the ``decretion'' disk around a Be star whose origin is thought to be related to rapid stellar rotation \citep[e.g.,][]{Coe2000,Reig2011}.
In most HMXBs, the accreting object is a highly-magnetized neutron star and pulsations are seen on the spin period of the neutron star as accreting matter is channeled onto its magnetic poles. Modulation is also typically seen on the system's orbital period due to effects including variable accretion rates for eccentric orbits, and eclipses by the primary, particularly for supergiant systems. On longer timescales, some systems exhibit periodic modulation on timescales longer than the orbital period - superorbital periods \citep[e.g.,][]{Corbet2013,Townsend2020}, and non-periodic long-term variability can also occur. For Be star systems, the decretion disk can dissipate \citep[e.g.][]{Reig2011}, causing a significant decrease in X-ray emission. Conversely, when the decretion disk is more extensive, very large ``Type II" outbursts can occur, which are not necessarily modulated on the orbital period as are ``Type I" outbursts \citep[e.g.,][]{Kuehnel2015}.

\src\ is a transient X-ray pulsar discovered with the Rossi X-ray Timing Explorer (RXTE) Proportional Counter Array (PCA) with a pulse period of 7.8 s \citep{Markwardt2004, Halpern2004} and it has a probable near-infrared counterpart \citep{Halpern2007} that is heavily-reddened, E(R - K) $>$ 10. RXTE PCA Galactic Bulge source monitoring found outbursts with a recurrence time of \sqig246 days lasting for \sqig7 days \citep{Markwardt2009}. A cyclotron resonance scattering feature (CRSF) was reported at \sqig18 keV by \citet{Roy2011} from RXTE PCA data, and \citet{Shtykovsky2019}
describe a \sqig15 keV CRSF found using the Nuclear Spectroscopic Telescope Array (NuSTAR). From the CRSF \citet{Shtykovsky2019} determined a magnetic field of 1.7 $\times$ 10$^{12}$ G. The X-ray spectrum also shows an iron line near 6.4 keV \citep{Roy2011,Shtykovsky2019}.
The distance of the source is uncertain, and estimates range from 4.5 to 18 kpc \citep{Halpern2007, Sguera2020} depending on the spectral type of the companion, which is not yet known. 
\citet{Halpern2007} noted that while XTE J1829-098 appeared most likely to be a Be star system, an OB supergiant or a red giant could not be excluded.
\citet{Christodoulou2022}
concluded that \src\ could either be
a Be star system at a distance of 4.5\,kpc, or a Supergiant Fast X-ray Transient located at 18\,kpc.
However, the supergiant classification would require the neutron star to have a mass of \sqig2.6\,\msun.

In this paper, we present X-ray observations of \src\ spanning almost 20 years obtained with the RXTE PCA, the Neil Gehrels Swift Observatory Burst Alert Telescope (BAT) and the Monitor of All-sky X-ray Image (MAXI). 
We show that, after initially being detected in an active state, it entered an extended period of quiescence for several years before becoming active again.
We find that the orbital-phase averaged properties remained similar, with periodic outbursts limited to a short phase range. We compare \src\ with the small number of Be star HMXBs and high-mass gamma-ray binaries that exhibit flares confined to a very short binary-phase range.
Unless otherwise noted, all uncertainties are given at the 1$\sigma$ level.

\section{Data and Analysis}

\subsection{PCA Galactic Plane Scans}

The PCA \citep{Jahoda2006} on board RXTE \citep{Bradt1993} covered an energy range 2--60 keV with a collimated detector.
Monitoring of regions around the Galactic center was carried out by
\citet{Swank2001} and \citet{Markwardt2006} who performed sets of scans to obtain source fluxes. The time resolution of each observation is given by how long a source was within the FOV of the PCA as it scanned over the source location, and this was typically 20 s.
Here we use a light curve from the RXTE PCA scans\footnote{https://asd.gsfc.nasa.gov/Craig.Markwardt/galscan/} with an energy range of 2--10 keV \citep{Markwardt2006} that covers from MJD 53,147 to 55,869 (2004-05-22 to 2011-11-04). This is approximately 1.5 years longer than the light curve used by \citet{Markwardt2009}.
Observations were generally obtained twice per week, and
the median spacing between the center of individual PCA observations is 3.66 days, with a minimum spacing of 1.96 days, and a maximum spacing of 101.66 days.

\subsection{BAT Observations}

The \Swift\ BAT is a hard X-ray telescope that uses
a coded mask to provide a wide field of view \citep{Barthelmy2005}.
Here we use light curves from the \Swift\ BAT transient monitor \citep{Krimm2013}, which
cover 15--50\,keV. 
In this energy range, the Crab gives a count rate of 0.22 \ctscm2s.
While transient monitor light curves
are available with time resolutions of both \Swift\ pointing durations 
(``orbital light curves''), and also daily averages, we used only the orbital light curve for \src.

The observation durations of the BAT orbital light curve
range from 62 to 2664 s and the mean duration is 606 s.
The median spacing between the center of individual BAT observations is 0.062 days (5564 s), with a minimum spacing of  52 s, and a maximum spacing of 38.8 days.
The BAT light curve of \src\ used here covers a time range
of MJD 53,416 to 60,359 (2005-02-15 to 2024-02-19).
We note that the transient monitor BAT light curve of \src\ that was available
for download at the time of writing only extended back to MJD 59,286 (2021-03-13). For this reason earlier data that had been
obtained from a previous download, which did include times between MJD 53,416 to 59,359,
was merged with a more recent light curve.

Our analysis was similar to that described in \citet{Corbet2013}.
We only use data for which the quality flag (``DATA\_FLAG'') was
0, indicating good quality. In addition, as in our previous work, we
removed points with very low fluxes which also had implausibly small uncertainties.
While changes to the transient light curve processing have reduced this problem,
some anomalous low flux points were still present before filtering.

\subsection{MAXI Observations}

MAXI, on board the International Space Station (ISS), carries two types of X-ray cameras: Gas Slit Camera (GSC) and Solid-state Slit Camera (SSC) \citep{Matsuoka2009, Mihara2011}. We used data from the GSC
covering an energy range of 2 -- 30 keV. Two types of GSC data are available: automatically generated light curves available for download with time resolutions of one day and one ISS orbit, and ``on-demand'' light curves\footnote{http://maxi.riken.jp/mxondem/}. 
We examined both types of MAXI light curves
for \src, and found that the automatically generated light curves exhibit considerable modulation on the $\sim$72 day precession period of the ISS, but this is much reduced in the on-demand light curves. Because of this, we use the on-demand light curve, and this covers a time range of MJD 55,062 (2009-08-19) to MJD 60,362 (2024-02-22). We specified ``1 orbit'' (93 minutes) time resolution and the resulting light curve has exposures ranging from 81 to 6070\,s with a mean of 358\,s.
The median spacing between the center of individual MAXI observations is 0.064 days (5564 s), with a minimum spacing of 0.015 days (1293\,s), and a maximum spacing of 41.3 days.
We also extracted a MAXI on-demand light curve for the Crab in the 2--30 keV energy band with the same time resolution, and find that this has a mean of 3.75 \ctscm2s.

\subsubsection{Calculation of Power Spectra}
As part of our analyses we calculate power spectra based on
discrete Fourier Transforms to search for periodic modulation. 
Our procedures for this were similar to those we have previously used, and for more details see \citet{Corbet2013} and \citet{Corbet2017}. 
Because the individual observations from all-sky observations typically have large exposure variations, it is advantageous
to weight data points when calculating the Fourier transform to maximize the signal-to-noise level.
\citet{Scargle1989} notes that weighting data points in calculating power spectra
is equivalent to a weighted combination of individual data points in the time domain. 
Our weighting uses
a factor which depends on both the uncertainty on each point, and the excess variability of the light curve which is mathematically equivalent to the semi-weighted mean \citep{Cochran1937,Cochran1954}. 
The significance of a peak in a power spectrum is given as a false alarm
probability \citep[FAP;][]{Scargle1982} and period uncertainties are obtained via
the expression of \citet{Horne1986}. The calculation of the FAP depends
on the number of independent frequencies, which depends on the frequency resolution
which is not precisely defined for unevenly sampled data \citep[e.g.,][]{Koen1990}. However, we have previously found that the inverse of the light-curve length does provide a reasonable
approximation for the frequency resolution for light curves such as those obtained by the BAT \citep[e.g.,][]{Corbet2017}.

\section{Results}

\subsection{The Light Curves}

\begin{figure}
\includegraphics[width=14cm,angle=270]{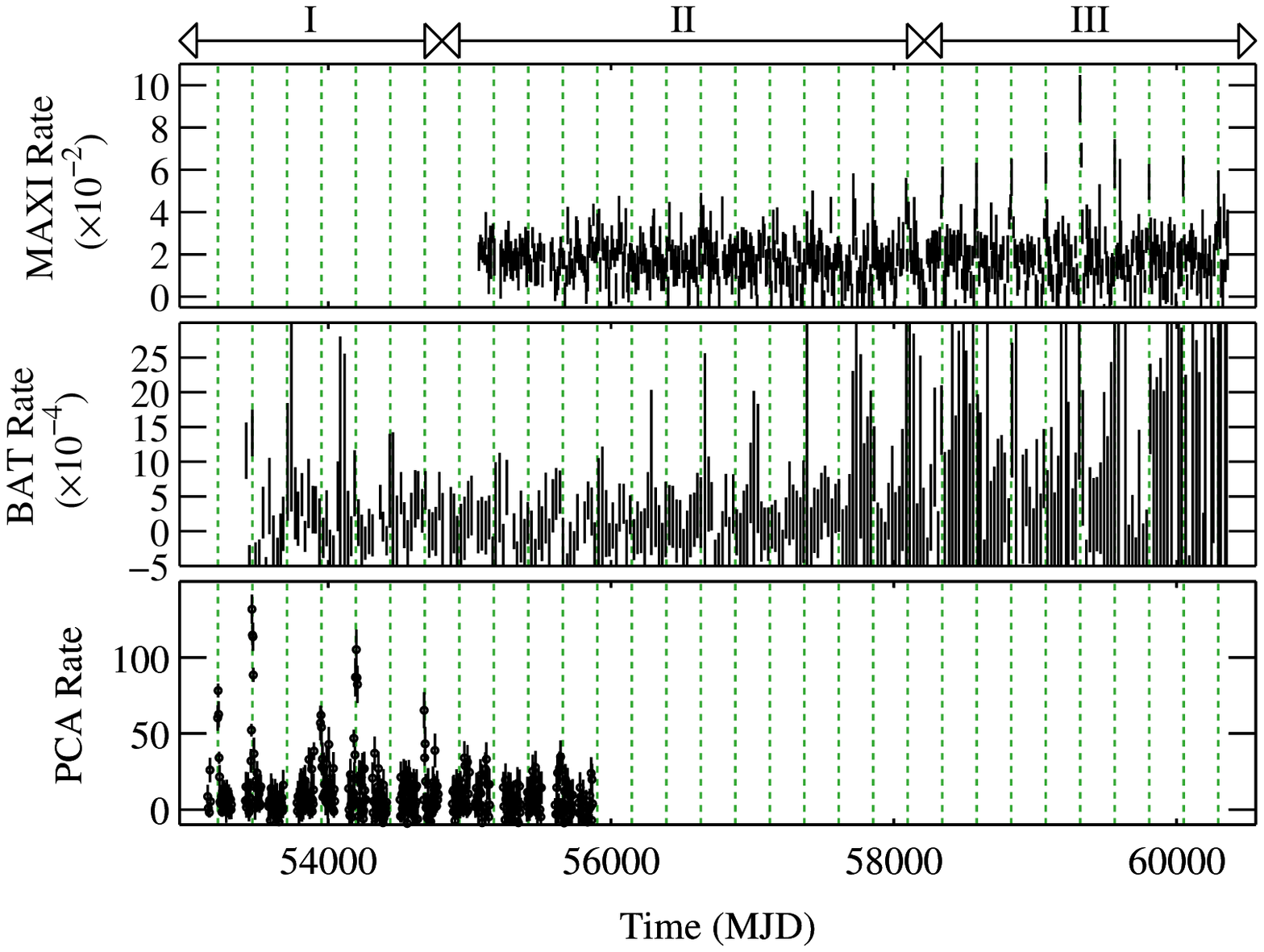}
\caption{
Light curves of \src\ obtained with the RXTE PCA (bottom), Swift BAT (middle) and MAXI (top). The
BAT and MAXI light curves were rebinned to 25 and 10 days respectively, while the PCA light curve was not rebinned from the original light curve.
The dashed green lines indicate the times of predicted source outbursts using our updated outburst ephemeris (Section \ref{section:profile_fits}, Table \ref{table:lc_joint_flare_fits}). The three activity epochs (``I'', ``II'', and ``III') are shown at the top of the plot.
During Epochs I and III outbursts are detected, while no strong outbursts are apparent during Epoch II.
}
\label{fig:multi_lc}
\end{figure}

The light curves from the PCA, BAT, and MAXI are plotted in Figure 
\ref{fig:multi_lc}. The BAT light curve is rebinned to a time resolution of 25 days, while the MAXI light curve has been rebinned to a time resolution of 10 days. 
In addition, in Figure \ref{fig:multi_lc} we only
retain BAT and MAXI data points that contain at least 10 input flux measurements. Without the rebinning and a  minimum number of input measurements, the BAT and MAXI light curves would be very noisy and flux variability would not be visible. 
The rebinned light curves are only used for plotting here, and we do not use them in subsequent data analysis. 
The times of predicted flares are marked with dashed green vertical lines. These come from our refined outburst ephemeris derived below in Section \ref{section:profile_fits}, but would be visually indistinguishable from outburst times using the earlier ephemeris of \citet{Nakajima2021}.

The flares from \src\ first reported from the PCA \citep{Markwardt2004,Markwardt2009} are clearly visible in the light curve and five flares are detected. Two predicted outbursts, centered at \sqig MJD 53,708 and \sqig 54,440 would have occurred during gaps in the PCA coverage. However, the last outburst that is clearly seen is at \sqig MJD 54,684. During the remainder of the light curve, the three predicted outbursts that do not fall into observing gaps are not significantly detected.

The signal-to-noise ratio of the BAT light curve is lower, even after rebinning, which hinders easy detection of individual outbursts during this interval.
We also note that the signal-to-noise of the BAT measurements has decreased with time. 
However, MAXI provides more sensitive measurements of the flux of \src. MAXI observations commenced on MJD 55,062, after the last PCA flare detection. The first report of a flare from \src\ with MAXI occurred between 2018-08-05 (MJD 58,335) to 2018-08-08 (MJD 58,338) \citep{Nakajima2018}. This outburst can be directly seen in the plot of the MAXI light curve (Figure \ref{fig:multi_lc}).
\citet{Nakajima2018} reported that the time of the outburst was consistent with both a \sqig246 day period \citep{Markwardt2009}
or a \sqig1.3 yr period \citep{Halpern2007}.
Subsequent to this outburst, additional flares at intervals of \sqig243 days are directly seen in the MAXI light curve.
Thus, there is an interval during which \src\ ceased to exhibit strong periodic flares and then subsequently resumed this behavior. 
Since the time of the transitions between the states is only shown by the flares, which are short in duration and separated by \sqig244 days, for our further analysis we select times which are intermediate between predicted flare times to separate activity states. We term the first activity state ``Epoch I'' and for the end of this, and the start of the
phase without strong outbursts we select MJD 54,805 (2008-12-05). We term this quiescent interval ``Epoch II'' and we use MJD 58,215 (2018-04-07) for the end of this. This is also the start of the second active state which we term ``Epoch III''.

\subsection{Folded X-ray Light Curves}

\begin{figure}
\includegraphics[width=14cm,angle=270]{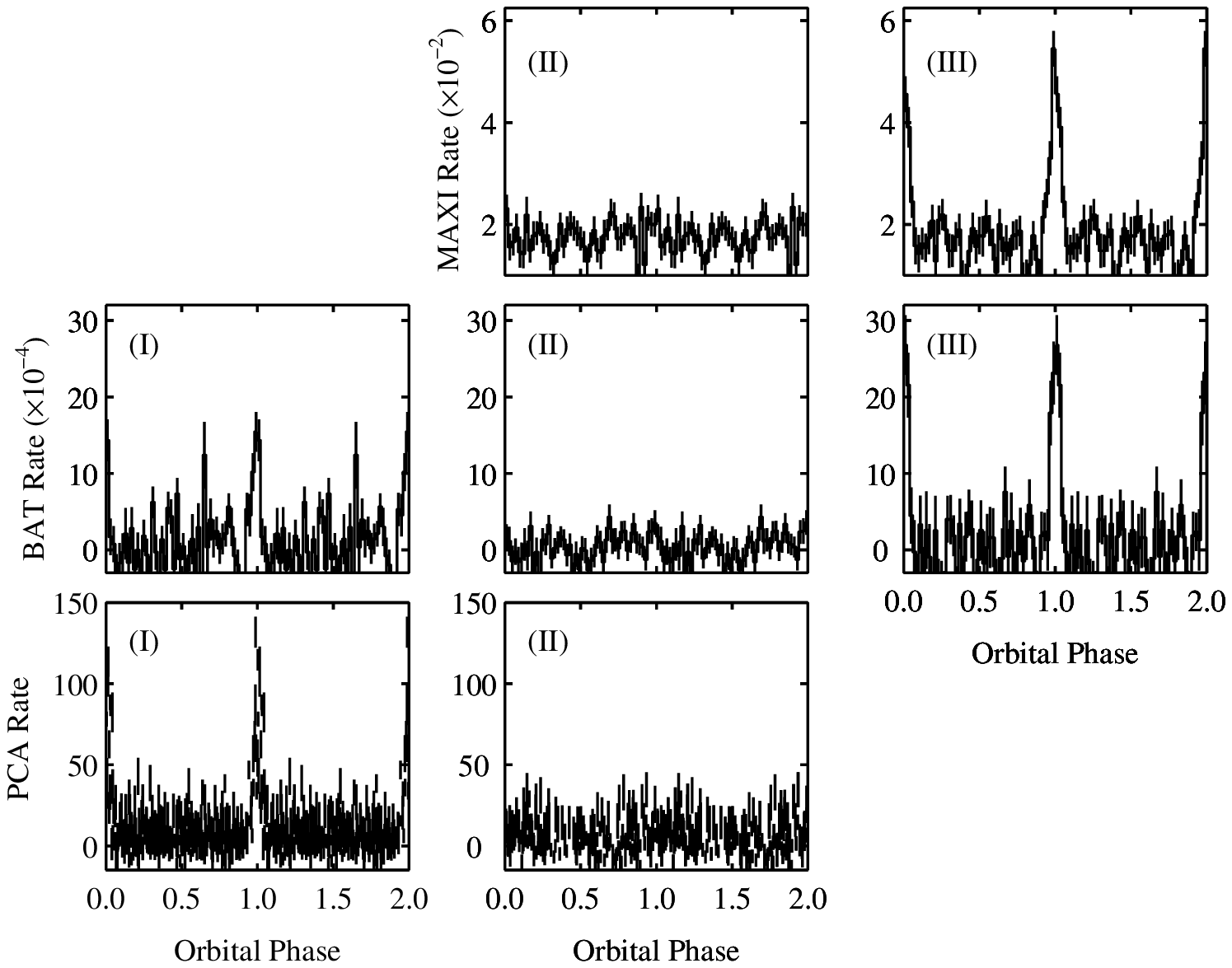}    
\caption{
Light curves of \src\ from RXTE PCA (bottom), Swift BAT (middle) and MAXI (top) observations folded on the orbital period. The BAT and MAXI light curves use 50 bins to cover one cycle, while the PCA light curve is not binned from the original light curve.
Data are selected for the three Epochs marked in Figure \ref{fig:multi_lc}.
}
\label{fig:all_fold}
\end{figure}

The light curves from all three instruments folded on the orbital period for the three Epochs separately are shown in Figure \ref{fig:all_fold}.
These again use our refined ephemeris derived below in Section \ref{section:profile_fits}, but would be similar using the previous ephemeris of \citet{Nakajima2021}.
The folded BAT and MAXI light curves both use 50 phase bins, while the PCA light curve is not phase-binned because the count rate uncertainties are smaller, and the observation cadence is also lower.
For Epoch I, the folded PCA light curve very clearly shows the sharp flares at phase zero. The folded BAT light curve for Epoch I, although noisier, does also show a sharp peak at that phase.
For Epoch II, as expected, none of the three instruments shows any sign of a significant outburst at phase zero. For Epoch III, for which only BAT and MAXI observations are available, these two instruments both show sharp outbursts at phase zero. For the BAT, the mean outburst profile shows a stronger maximum than during Epoch I. However, we note that the average profiles come from a relatively small number of outbursts, six for Epoch I and nine for Epoch III.

\subsection{Power Spectrum Analysis of the Active and Inactive States}
\label{section:power_spectra}

\begin{figure}
\includegraphics[width=14cm,angle=0]{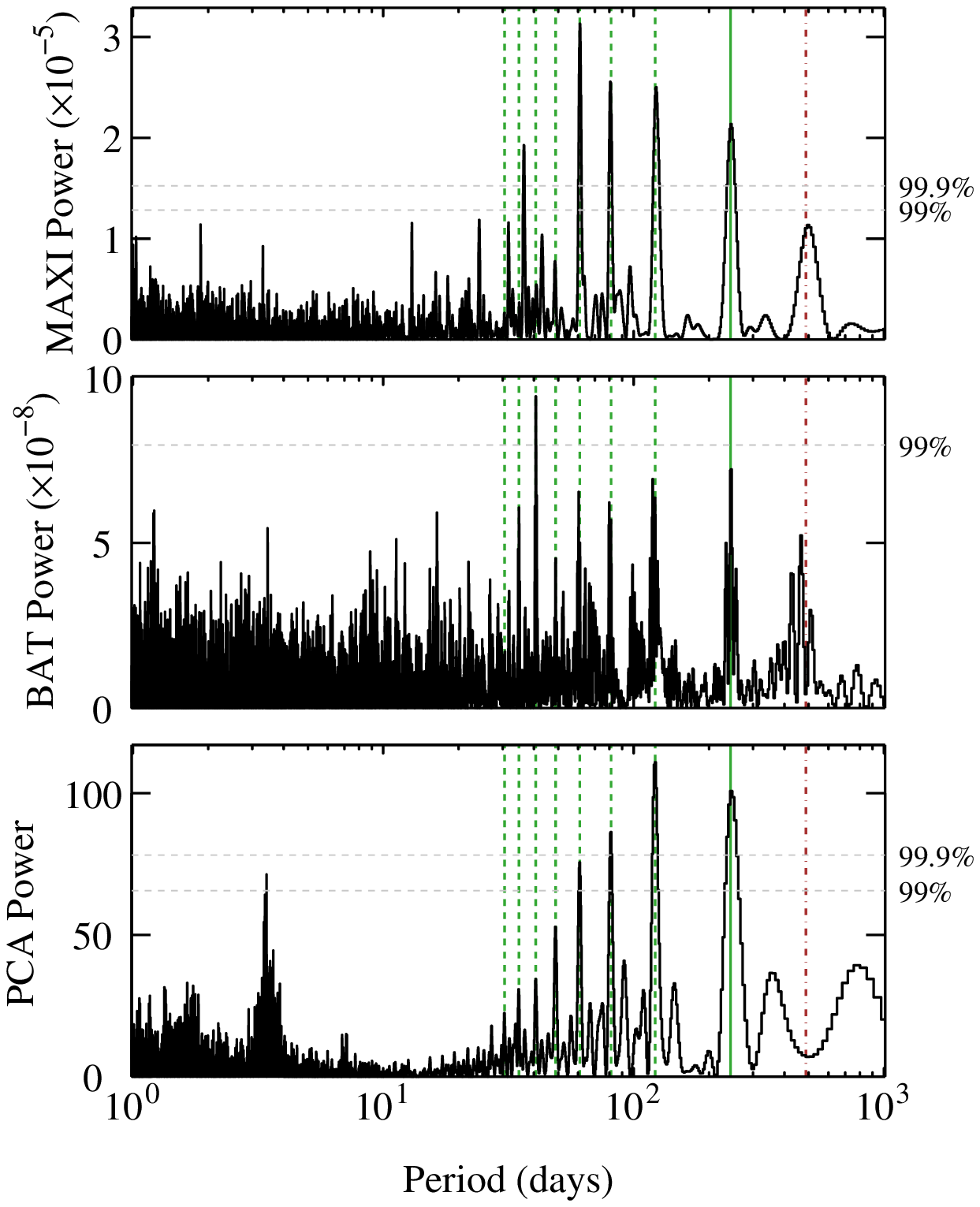}
\caption{
Power spectra of light curves of \src\ during Epoch I and III  obtained with the RXTE PCA (bottom), Swift BAT (middle) and MAXI (top). The solid green line shows the orbital period, and the dashed green lines show the harmonics of this. The dot-dashed red line marks twice the orbital period. The peak at \sqig 3.5 days in the power spectrum of the PCA light curve is due to the observing cadence.
}
\label{fig:sel_power}
\end{figure}

\begin{figure}
\includegraphics[width=14cm,angle=0]{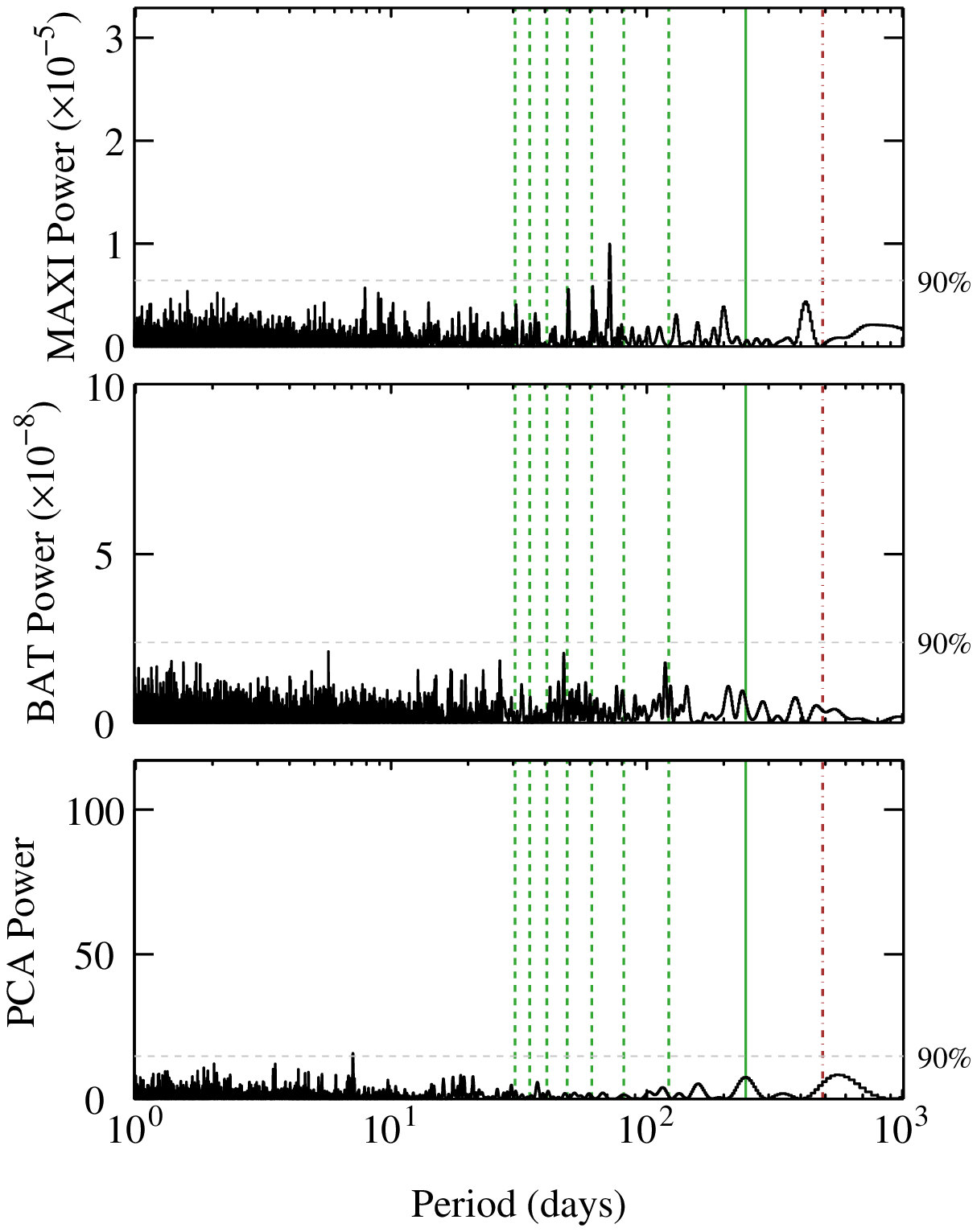}
\caption{
Power spectra of light curves of \src\ during Epoch II  obtained with the RXTE PCA (bottom), Swift BAT (middle) and MAXI (top). 
The solid green line shows the orbital period, and the dashed green lines show the harmonics of this.
The strongest peak in the MAXI power spectrum corresponds to the precession period of the ISS.
}
\label{fig:anti_sel_power}
\end{figure}

We calculated power spectra of the light curves for (i) the active intervals Epoch I and Epoch III together, and (ii)  the quiescent interval Epoch II. 
These are shown in Figures \ref{fig:sel_power} and \ref{fig:anti_sel_power} respectively.
For the power spectra of the PCA and MAXI light curves during Epochs I and III (Figure \ref{fig:sel_power} bottom and top respectively) strong peaks are clearly seen at the fundamental of the orbital period and also many harmonics of this. The presence of many peaks is consistent with the highly non-sinusoidal modulation of the light curves.
While the power spectrum of the BAT light curve does show peaks near the orbital period and its harmonics, 
it shows significant aliasing due to the gap between Epoch I and Epoch III. The periods of the peaks harmonically related to the orbital period are listed for each instrument in Table \ref{table:power_spectrum_peaks}. {\mybf From a zero-order polynomial fit to the} period measurements from all harmonics of all three instruments we derive an orbital period of 243.83 $\pm$  0.35 days.
The power spectrum of the MAXI light curve shows a modest peak near twice this period and we discuss the possibility of modulation at twice the \sqig244 day period in Appendix \ref{section:double_period}.
For Epoch II (Figure \ref{fig:anti_sel_power})  no significant modulation is found from the power spectra from either instrument. Constraints on modulation during Epoch II are also discussed below in Section \ref{section:flare_constraints}.

\begin{deluxetable}{lccc}
\tablecolumns{8}
\tabletypesize{\small}
\tablewidth{0pc}
\tablecaption{Peaks in Power Spectra of \src}
\tablehead{
\colhead{Harmonic} & \colhead{PCA I} &  \colhead{BAT I+III} & \colhead{MAXI III} \\
\colhead{} & \colhead{(days)} &  \colhead{(days)} & \colhead{(days)}  \\
}
\startdata
1 & 246.04 $\pm$ 6.4                   & 244.71 $\pm$ 1.14  & 244.8 $\pm$  3.7  \\
2 & 121.6 $\pm$ 1.4 (243.3 $\pm$ 2.8) & -& 122.9 $\pm$  1.0    (245.9  $\pm$ 1.9)   \\
3 & 81.0 $\pm$ 0.7 (243.1 $\pm$ 2.2) & -& 80.85 $\pm$  0.43   (242.5  $\pm$ 1.3) \\
4 & 60.8 $\pm$ 0.4 (243.3 $\pm$ 1.5) & -& 61.04  $\pm$ 0.23   (244.17  $\pm$  0.91) \\
5 & 48.8 $\pm$ 0.3 (243.9 $\pm$ 1.3) & - & - \\
6 & 40.65 $\pm$ 0.11 (243.9 $\pm$ 0.7) & - & - \\
7 & 34.76 $\pm$ 0.15 (243.3 $\pm$ 1.0) & - & - \\
8 & 30.45 $\pm$ 0.13 (243.6 $\pm$ 1.0) & - & - \\
\tableline
Instrument Mean & 243.66 $\pm$ 0.43 & 244.71 $\pm$  1.14 & 243.95 $\pm$ 0.68  \\
\tableline
Mean of All Peaks & 243.83 $\pm$  0.35 & & \\
\enddata
\tablecomments{
Numbers in parentheses are the implied fundamental period from each harmonic.
Mean is taken for the fundamental periods.
No period is given if the harmonic is at noise level, or an alias peak is stronger than the expected peak.
}
\label{table:power_spectrum_peaks}
\end{deluxetable}

\subsection{Characterizing the Periodic Flares and Refining the Orbital Period}
\label{section:profile_fits}

The folded light curves (Figure \ref{fig:all_fold}) show sharp flares confined to a narrow phase range. To determine the orbital period and other parameters we fitted a profile consisting of a periodic sharp flare together with a fixed flux level between these for all light curves. We note that these fits were made to the original light curves, and {\em not} the folded light curves.
This triangular profile is defined in terms of time of orbital period, the maximum rate ($R_{max}$) above background, the epoch of maximum rate, which defines phase zero, the phase of the start of the flare ($\phi_{\rm start}$),
the phase of the end of the flare ($\phi_{\rm end}$), and the 
background rate outside eclipse ($R_{\rm base}$).
Similarly to the computation of the power spectra, the data points were weighted by their errors.

\begin{equation}
\label{RateEquation}
  R = R_{\rm base} +
  \left\{
    \begin{array}{ll}
      0, 
& \phi_{\rm end} < \phi < \phi_{\rm start} \\
      R_{\rm max}  (\frac { (\phi -  \phi_{\rm start})}{(1 - \phi_{\rm start})})
   , 
& \phi \geq \phi_{\rm start}  \\

       R_{\rm max}  (1 - \frac {\phi}{\phi_{\rm end}}) , 
&  \phi \leq \phi_{\rm end} \\

    \end{array}
  \right.
\end{equation}

We note that the function is potentially asymmetric, as the rise duration from $R_{\rm base}$ to $R_{\rm max}$ (1 - $\phi_{\rm start}$) and the decay duration from $R_{\rm max}$ to $R_{\rm base}$
($\phi_{\rm end}$) may be different.

We first fitted the light curves from each instrument independently.
In addition, since the BAT folded profile from Epoch I shows an apparently lower amplitude than the folded profile from Epoch III, we fit the two epochs separately for the BAT.
The results from this are given in Table \ref{table:lc_flare_fits}.
While the PCA and MAXI light curves have non-zero fluxes outside of the flares, we ascribe this to incomplete background subtraction rather than emission from \src\ itself.

\begin{deluxetable}{lccccc}
\tablecolumns{8}
\tabletypesize{\small}
\tablewidth{0pc}
\tablecaption{Separate Fits to Periodic Outbursts of \src}
\tablehead{
\colhead{Parameter} & \colhead{PCA I} & \colhead{BAT I+III} & \colhead{BAT I}& \colhead{BAT III}& \colhead{MAXI III}  \\
}
\startdata
Orbital Period (days) & 243.61 $\pm$ 0.08 & 244.22 $\pm$ 0.09 & 243.69 $\pm$ 0.18 & 243.95 $\pm$ 0.36 & 243.65 $\pm$ 0.21   \\
Epoch Flare Maximum (MJD) & 58,338.1 $\pm$ 1.0 & 58,343.4 $\pm$ 1.5 & 58,333.9 $\pm$ 2.1 & 58,342.6 $\pm$ 1.7 & 58,343.6 $\pm$ 1.1  \\ 
Phase Flare Start ($\phi_{\rm start}$) & 0.919 $\pm$ 0.010 & 0.929 $\pm$ 0.011 & 0.949 $\pm$ 0.021 & 0.934 $\pm$ 0.013 & 0.921 $\pm$ 0.008   \\
Phase Flare End ($\phi_{\rm end}$)& 0.057 $\pm$ 0.006 & 0.059 $\pm$ 0.011 & 0.044 $\pm$ 0.015 & 0.069 $\pm$ 0.012 & 0.082 $\pm$ 0.009   \\
Maximum Count Rate ($R_{\rm max}$) & 71.6 $\pm$ 3.9 & 19.3 $\pm$ 1.7 & 16.9 $\pm$ 2.1 & 28.2 $\pm$ 3.6 & 3.77 $\pm$ 0.24  \\
Base Count Rate ($R_{\rm base}$) & 3.0 $\pm$ 0.6 & 0.1 $\pm$ 0.3 & 0.4 $\pm$ 0.3 & -0.3 $\pm$ 0.5 & 1.59 $\pm$ 0.05  \\
$\chi^2_{\nu}$ & 3.4 & 1.05 & 1.07  & 1.04 & 1.05  \\
\tableline
Flare Duration FWZI (phase)$^1$ & 0.138 $\pm$ 0.012 & 0.130 $\pm$ 0.014 & 0.095 $\pm$ 0.025 & 0.135 $\pm$ 0.018 & 0.161 $\pm$ 0.012\\
Flare Duration FWZI (days) & 33.6 $\pm$  2.9 & 31.7 $\pm$  3.4 & 22.9  $\pm$ 6.1 & 32.9 $\pm$  4.4  & 39.1 $\pm$  2.9\\
\enddata
\tablecomments{Data from the quiescent interval (Epoch II) is excluded.
Parameters below the line are derived from those above it. ``BAT I'' is a fit to only data from Epoch I, and ``BAT III'' is a fit to only data from Epoch III. PCA observations do not cover Epoch III, and MAXI observations do not cover Epoch I.
BAT count rates are in units of 10$^{-4}$ and MAXI count rates are in units of 10$^{-2}$.
The fitted function is defined in Equation \ref{RateEquation}. Note that $R_{\rm max}$ is measured with respect to $R_{\rm base}$ and not zero.
$^1$Flare Duration FWZI (phase) =  $1 - \phi_{\rm start} + \phi_{\rm end}$.
}
\label{table:lc_flare_fits}
\end{deluxetable}

\begin{figure}
\includegraphics[width=14cm,angle=0]{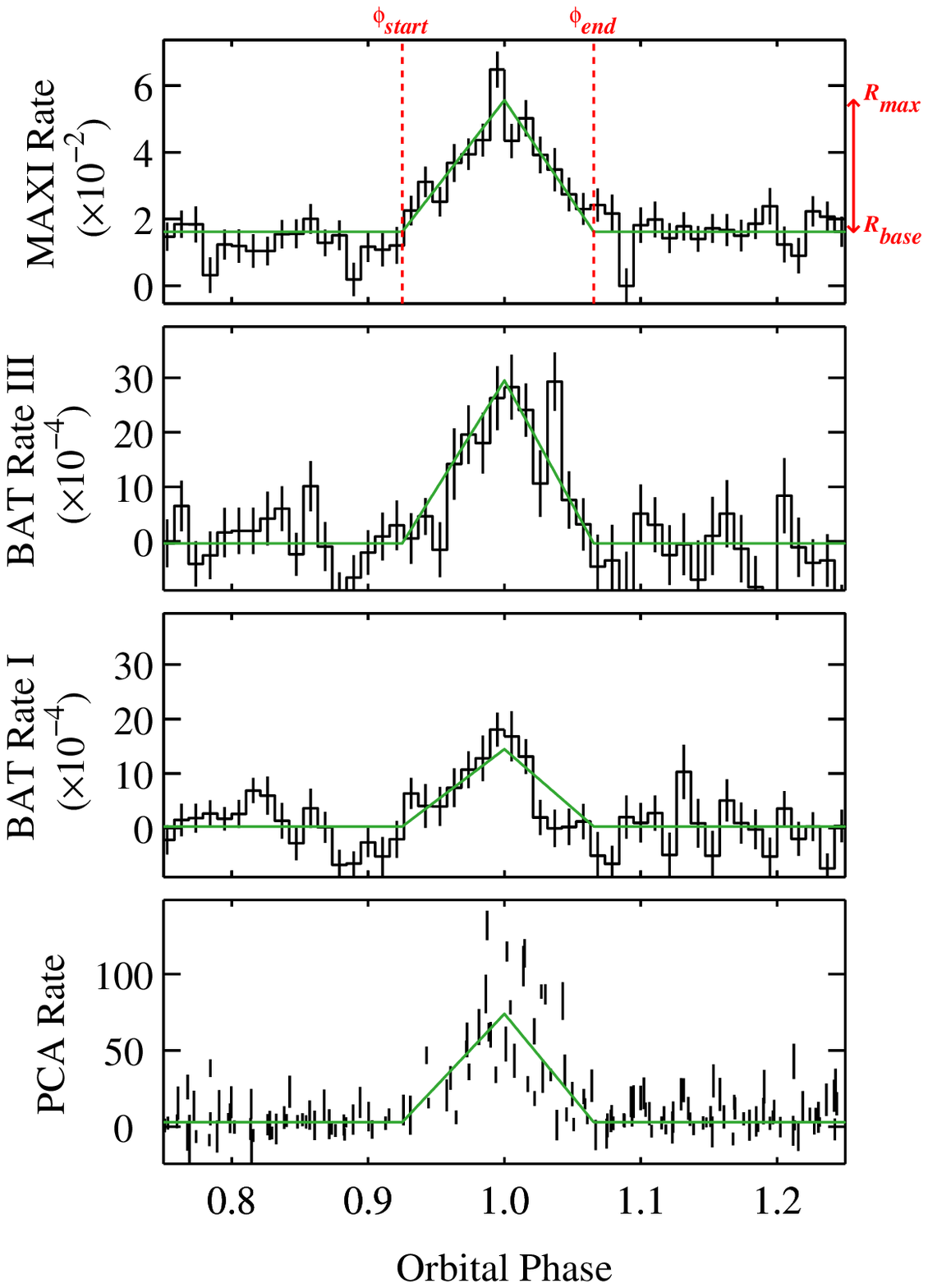}    
\caption{
Light curves of \src\ from RXTE PCA, Swift BAT and MAXI observations folded on the orbital period using only data from Epochs I and III.
The BAT and MAXI light curves were folded into 95 bins for one cycle, but only phases near periastron passage (0.75 to 1.25) are plotted.
The green lines mark the triangular profile from Equation \ref{RateEquation} fitted to the unfolded light curves from those epochs. Parameters are given in Table \ref{table:lc_joint_flare_fits}.
}
\label{fig:four_flare_fit}
\end{figure}

We found that the overall profiles obtained from each instrument are
broadly comparable. Because of this, we then simultaneously fit all three light curves to both provide the longest possible time base and improve the parameter determinations. 
This makes use of the different advantages of the three light curves: 
The PCA provides the highest signal-to-noise ratio for individual flux measurements, but the observations have a very low cadence and the light curve only covers Epoch I and the start of Epoch II. The BAT observations have higher cadence and were obtained over all three Epochs, although they have lower signal-to-noise, particularly for the more recent data. The MAXI light curve has similar cadence to the BAT, but higher signal-to-noise for \src, and provides observations over Epoch II and Epoch III.
For this joint fit, the orbital period, epoch of maximum flux, $\phi_{\rm start}$, and $\phi_{\rm end}$ were the same for all instruments, but $R_{max}$ and $R_{base}$ were allowed to float for each instrument.
The results from this are given in Table \ref{table:lc_joint_flare_fits}
and the fits and folded light curves are plotted in Figure \ref{fig:four_flare_fit}. We find an orbital period of 243.95 $\pm$ 0.04 days. This is consistent with both
the period determined from our power spectra of 243.83 $\pm$ 0.35 days (Section \ref{section:power_spectra}), and the period of 243.66 $\pm$ 3.07 days reported by \citet{Nakajima2021}, but with a much smaller uncertainty.
Our epoch of maximum flux, MJD 58,342.3 $\pm$ 0.7, is also consistent with that reported by \citet{Nakajima2021} of MJD 58,341.7 (no error reported). {\mybf Our ephemeris is consistent with, but provides tighter constraints than, that recently reported by \citet{Nakajima2024} from MAXI data alone of epoch of maximum flux = MJD 58,342.1 + n $\times$ (243.8 $\pm$ 1.5 ).}
The derived mean profile does not show any statistically significant asymmetry as the difference between the rise and decay times to and from peak is 2.3 $\pm$ 1.7 days.

\begin{figure}
\includegraphics[width=14cm,angle=270]{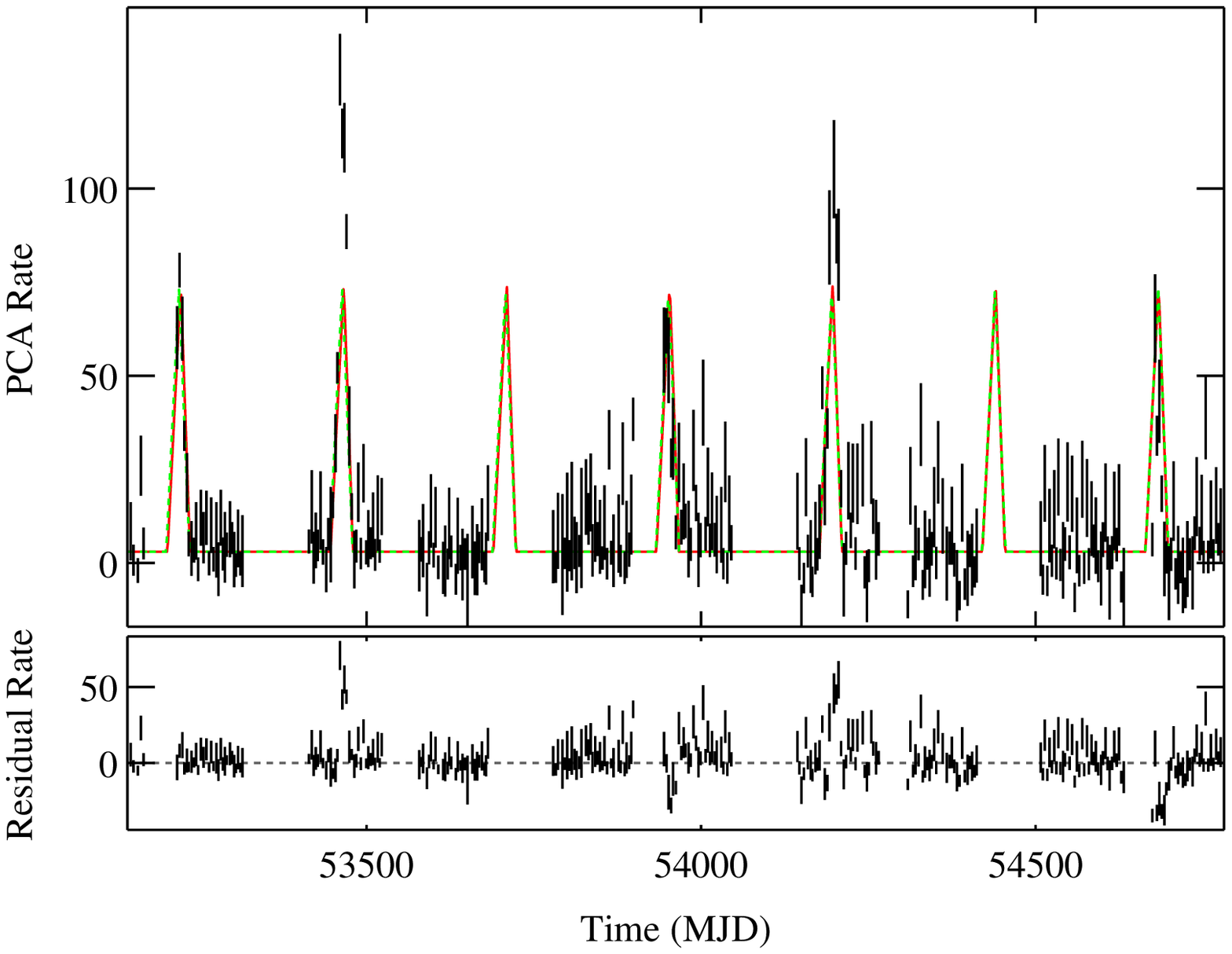}    
\caption{
{\mybf Top:} The PCA light curve of \src\ for Epoch I. The red line shows the fit to the outbursts from the PCA alone (Table \ref{table:lc_flare_fits}) and the dashed green line shows the results from simultaneously fitting all three light curves (Table \ref{table:lc_joint_flare_fits}).
{\mybf Bottom: the residuals from the simultaneous fit.}
}
\label{fig:pca_show_fits}
\end{figure}

While comparing fluxes between different instruments covering different energy ranges is challenging, we note that the $R_{\rm max}$ obtained from Epoch III with MAXI corresponds to 10.5 $\pm$ 0.7 mCrab, while $R_{\rm max}$ obtained for this Epoch with the BAT corresponds to 13.6 $\pm$ 1.2 mCrab, and so are rather similar. For Epoch I, the BAT peak rate is equivalent to 6.5 $\pm$ 0.8 mCrab.
For the PCA, we take the statement from \citet{Markwardt2004} that the count rate on 2004-07-30.9 (60.21\,\ctss) corresponds to \sqig7 mCrab. This implies that $R_{\rm max}$ for the PCA (Epoch I) is equivalent to 8.2 $\pm$ 0.4 mCrab. This may lend support to the mean outburst intensity during Epoch I being lower than Epoch III. But, even if this is the case, cycle-to-cycle variability might be the cause rather than an overall difference between the two active Epochs.
Evidence for cycle-to-cycle variability comes from the higher precision flux measurements from the PCA. 
In the folded light curve (Figure \ref{fig:four_flare_fit} bottom) there is 
clearly significant deviation from the mean profile. This can be further seen when the fits are plotted over the PCA light curve (Figure \ref{fig:pca_show_fits}). In this figure we also plot the fitted profile obtained from the separate fit to the PCA data (Table \ref{table:lc_flare_fits}). This essentially overlaps the joint fit, giving confidence that it is reasonable to  simultaneously fit all the light curves.

{\mybf To compare the maximum fluxes ($R_{max}$) during the two active epochs with previous observations we use a conversion factor of 1\,mCrab = 2.4$\times10^{-11}$\ergscm2s\ (2 -- 10 keV) as also employed by \citet{Halpern2007}. 
For Epoch I we derive a F$_X$ (2 - 10 keV) = 1.9 ($\pm0.1$)\, $\times10^{-10}$\,\ergscm2s, and for Epoch III we find F$_X$ (2 - 10 keV) = 2.3 ($\pm0.1$)\, $\times10^{-10}$\,\ergscm2s. These are broadly consistent with fluxes obtained from pointed observation near phase 0.0 (Table \ref{table:flux_pulse_history}).
}

\begin{deluxetable}{lccccc}
\tablecolumns{8}
\tabletypesize{\small}
\tablewidth{0pc}
\tablecaption{Joint Fits to Periodic Outbursts of \src}
\tablehead{
\colhead{Parameter} & \colhead {All} & \colhead{PCA I} & \colhead{BAT I} & \colhead{BAT III} & \colhead{MAXI III}  \\
}
\startdata
Orbital Period (days) & 243.95 $\pm$ 0.04 &  &  & & \\
Epoch Flare Maximum (MJD) & 58,342.3 $\pm$ 0.7 &  &  & & \\
Phase Flare Start ($\phi_{\rm start}$) & 0.925 $\pm$ 0.006 &  &  & & \\
Phase Flare End ($\phi_{\rm end}$) & 0.066 $\pm$ 0.005 &  &  & & \\
Maximum Count Rate ($R_{\rm max}$) &  & 70.9 $\pm$ 3.6 & 14.2 $\pm$ 1.7 & 29.9 $\pm$ 2.7 & 3.95 $\pm$ 0.25  \\
Base Count Rate ($R_{\rm base}$) &  & 3.0 $\pm$ 0.6 & 0.3 $\pm$ 0.3 & -0.4 $\pm$ 0.5 & 1.61 $\pm$ 0.05  \\
$\chi^2_{\nu}$ & 1.05 &  &  & \\
\tableline
Flare Duration FWZI (phase)$^1$ & 0.140 $\pm$ 0.007 &  &  & & \\
Flare Duration FWZI (days) & 34.3 $\pm$ 1.7 &  &  &  & \\
Delta (Rise - Decay) (phase)$^2$ & 0.009 $\pm$ 0.007 &  &  & & \\
Delta (Rise - Decay) (days) & 2.3 $\pm$ 1.7 &  &  & & \\
\enddata
\tablecomments{Data from the quiescent interval (Epoch II) is excluded.
BAT I = fit to BAT light curve from Epoch I, BAT III = fit to BAT light curve from Epoch III.
Parameters below the line are derived from those above it. PCA observations do not cover Epoch III, and MAXI observations do not cover Epoch I.
BAT count rates are in units of 10$^{-4}$ and MAXI count rates are in units of 10$^{-2}$.
The fitted function is defined in Equation \ref{RateEquation}.
$^1$Flare Duration FWZI (phase) =  $1 - \phi_{\rm start} + \phi_{\rm end}$.
$^2$Delta (Rise - Decay) (phase) = $1 - \phi_{\rm start} - \phi_{\rm end}$.
}
\label{table:lc_joint_flare_fits}
\end{deluxetable}

\subsection{Previous Flux and Pulse Period Measurements}

\begin{figure}
\includegraphics[width=14cm,angle=270]{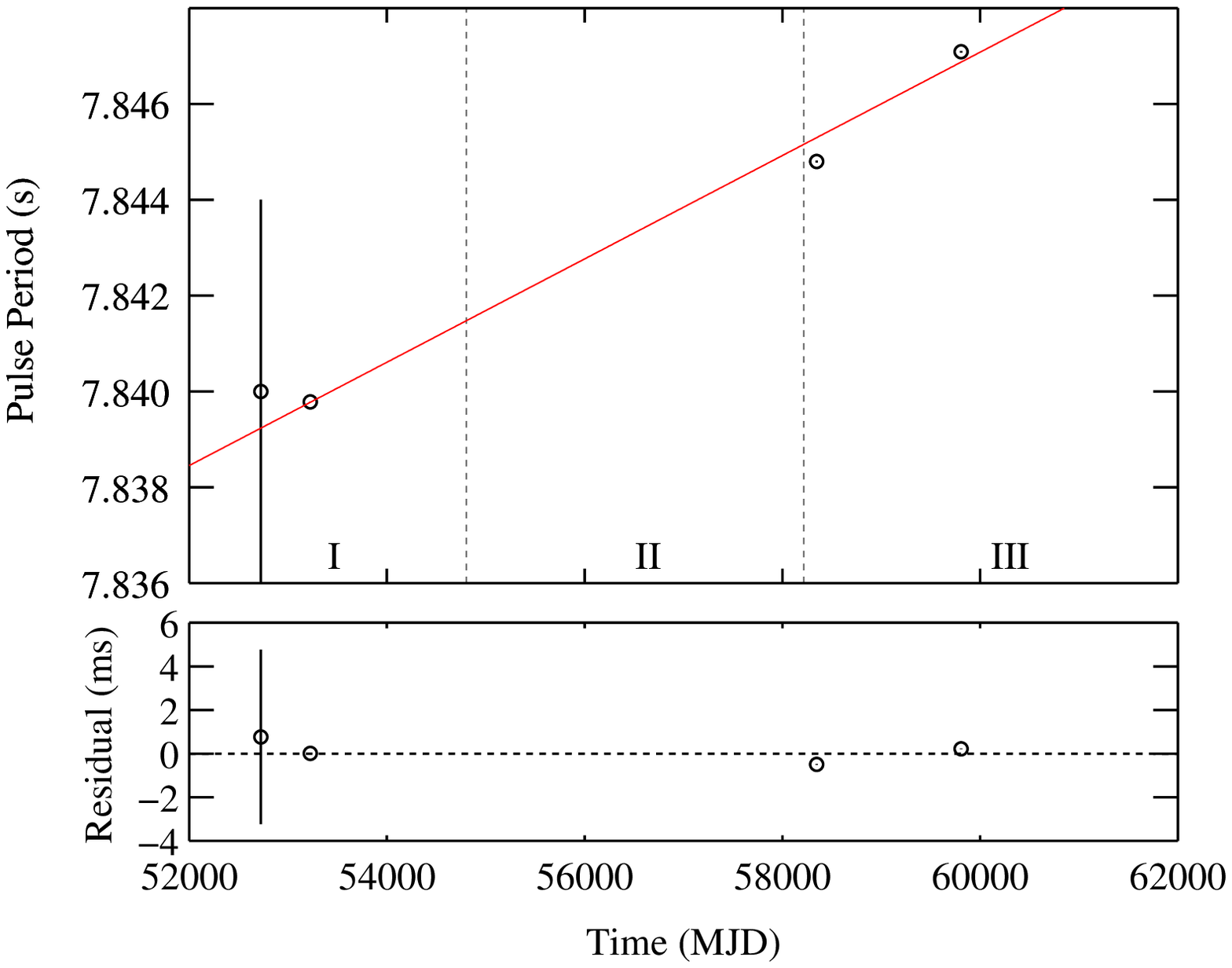}    
\caption{
Pulse period history of \src\ taken from Table \ref{table:flux_pulse_history}.
The RXTE PCA measurement from MJD 53,216 is not plotted because of its large
uncertainty. The vertical dashed lines indicate the divisions between the three Epochs.
}
\label{fig:pulse_history}
\end{figure}

To compare previous measurements of \src\ we compiled a list
of flux and pulse period measurements and these are given
in Table \ref{table:flux_pulse_history} along with the calculated phase and the Epoch.
The pulse periods are also plotted in Figure \ref{fig:pulse_history} and,
from the few measurements with small uncertainties, a linear fit gives a spin-down rate of 
1.25(5)\,$\times$\,10$^{-11}$\,s\,s$^{-1}$ 
$\implies$ \PPdot\,\sqig6.3$\times$10$^{11}$\,s (\sqig20,000 yr).
We caution that, in addition to the small number of pulse period measurements, they were obtained from the two active states Epoch I and Epoch III, with unknown behavior in between, which hinders drawing strong conclusions on the long-term spin period changes. These pulse period measurements have also not been corrected for orbital Doppler shifts, although \citet{Markwardt2009} reported that they found no evidence of orbital modulation of pulse arrival times in their PCA observations.
The flux measurements in Table \ref{table:flux_pulse_history} are also consistent with the 244 day period as high fluxes have only been detected near phase zero for this period.

\clearpage
\begin{deluxetable}{lcccccc}
\tablecolumns{8}
\tabletypesize{\small}
\tablewidth{0pc}
\tablecaption{Previous Pointed Observations of \src}
\tablehead{
\colhead{Mission} & \colhead{Observation Time} & \colhead{Epoch} & \colhead{Phase} &
\colhead{Flux (2 - 10 keV)} & \colhead{Pulse Period} & \colhead{Ref.} 
\\
\colhead{}  & \colhead{(MJD)} &\colhead{} &\colhead{} & \colhead{(\ergscm2s)} & \colhead{(s)} & \colhead{} \\
}
\startdata
XMM & 52,360	&	-	&	0.48		&	$<5\times10^{-14}$ & & (a) \\
XMM & 52,725		& -		 & 0.97		 & 	$8.2\times10^{-11}$ & 7.840 $\pm$ 0.004 & (a), (i) \\
XMM & 52,895		& -		 & 0.67		 & 	$<5\times10^{-14}$ & & (a) \\
RXTE & 53,147 - 53,166	 & I	 & 	0.70 - 0.78	 & $<4.8\times10^{-11}$ & & (a) \\
RXTE & 53,216		& I		 & 0.99		 & 	$1.7\times10^{-10}$ & 7.82 $\pm$ 0.05  & (a), (ii) \\
RXTE & 53,225		& I		 & 0.02		 & 	$1.0\times10^{-10}$ & 7.839783 $\pm$ 0.000006 & (a), (iii) \\
Chandra & 54,145	& I	& 0.79		 & 	$2.5\times10^{-14}$  & & (a) \\
Chandra & 54,244		 & I & 0.20		 & 	$2.8\times10^{-13}$  & & (a) \\
NuSTAR & 58,346 & III & 0.01 & $3.6\times10^{-10}$   &  7.84480 $\pm$ 0.00002 & (b), (iv) \\
NICER & 59,807 & III & 0.00 & $3.9\times10^{-10}$ & 7.847089 $\pm$ 0.000015 & (c), (v) \\ 
\enddata
\tablecomments{Fluxes based on: (a) \citet{Halpern2007} (2 - 10 keV fluxes), (b) \citet{Shtykovsky2019} (3 - 79 keV), and (c) \citet{Wolff2022} (0.5 - 10 keV).
Pulse periods from: (i) \citet{Halpern2007}, (ii) \citet{Markwardt2004}, (iii) \citet{Markwardt2009}, (iv) \citet{Shtykovsky2019}, and (v) \citet{Wolff2022}.
Pulse periods are also plotted in Figure \ref{fig:pulse_history}.
}
\label{table:flux_pulse_history}
\end{deluxetable}

\subsection{Constraints on Sharp Flares During Quiescent Epoch II}
\label{section:flare_constraints}

To obtain limits on sharp flares during Epoch II, we fitted
the same model as used for Epochs I and III, with all parameters frozen to the best fit values (Table \ref{table:lc_joint_flare_fits}) with the exception of $R_{\rm max}$
and $R_{\rm base}$. From this we obtained 
PCA: $R_{\rm max}$ = 2.4 $\pm$ 4.0, $R_{\rm base}$ = 2.5 $\pm$ 0.9; 
BAT: $R_{\rm max}$ = (2.8 $\pm$ 1.1) $\times$ 10$^{-4}$, $R_{\rm base}$ = (0.3 $\pm$ 0.2) $\times$ 10$^{-4}$; 
MAXI: $R_{\rm max}$ = (0.39 $\pm$ 0.18) $\times$ 10$^{-2}$, $R_{\rm base}$ = (1.73 $\pm$ 0.04) $\times$ 10$^{-2}$.
Thus, there is no significant detection ($\geq$3$\sigma$) of an outburst from any instrument during Epoch II.
The lack of detection of modulation in the power spectra during Epoch II (Section \ref{section:power_spectra}) also shows that there is no strong modulation with a broader profile.

\section{Discussion}

\subsection{Location of \src\ in the Spin/Orbital Period Diagram}

\begin{figure}
\includegraphics[width=14cm,angle=270]{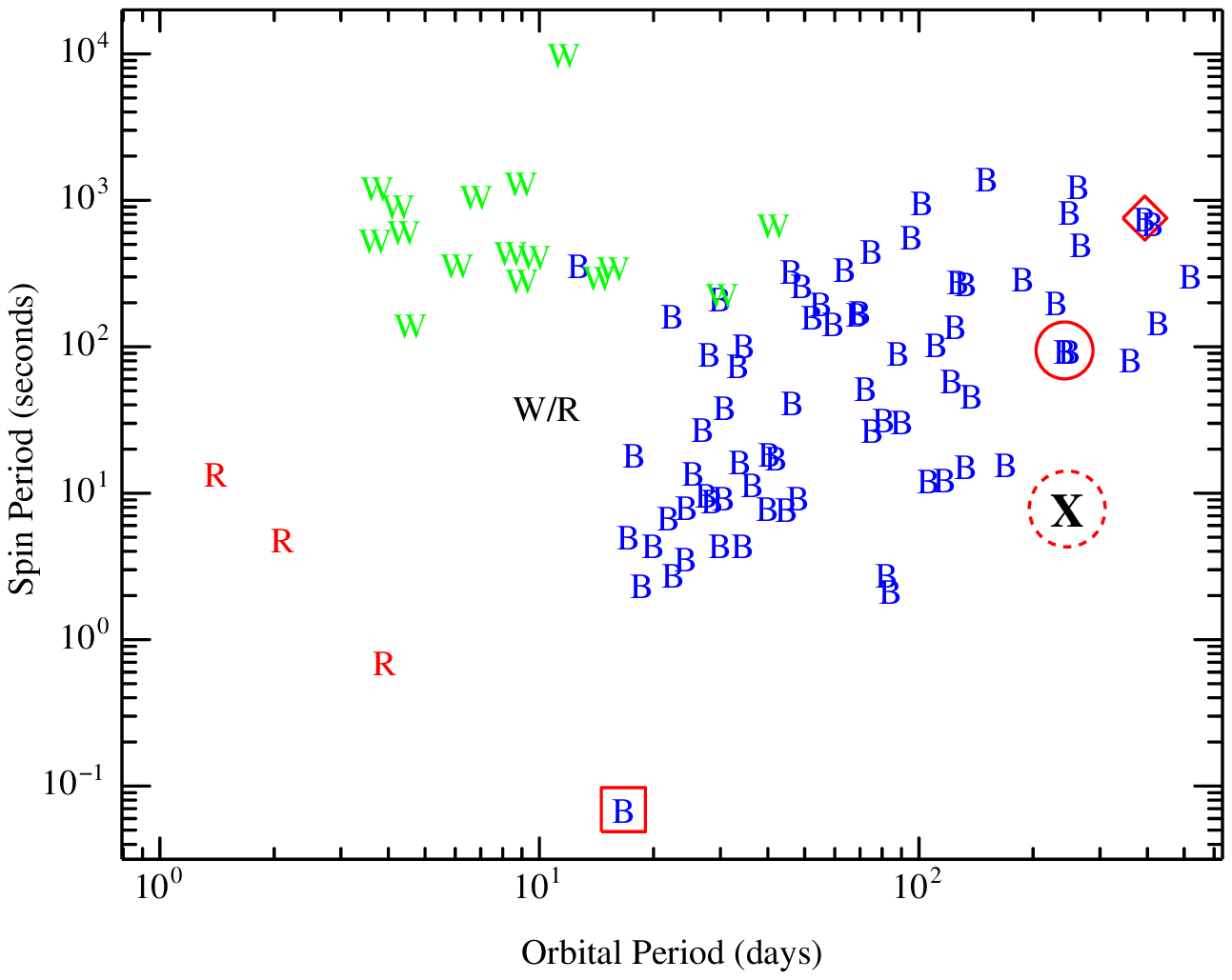}    
\caption{
Spin period vs. orbital period for high-mass X-ray binaries. The letters indicate the type of mass transfer believed to be taking place in each system. R = Roche-lobe overflow, W = wind accretion from a supergiant, B = accretion from a Be star decretion disk. The parameters of \src\ are marked with the dashed circled red X, those of GS 1843-02 with the solid red circle,  \a0538\ with the red box, and AX J0049.4-7323 with the red diamond. For system parameters see, e.g., \citet{Fortin2023,Neumann2023,Kim2023}.
}
\label{fig:stars}
\end{figure}

A useful diagnostic for HMXBs is the diagram showing neutron star spin period versus orbital period \citep{Corbet1986}
and we show this in Figure \ref{fig:stars}.
This diagram shows the separation of sources depending on their mode of accretion: Be star envelope, wind from a supergiant, and Roche-lobe overflow.
There is also a general correlation between spin period and orbital period for the Be star systems.
While the degree of correlation is not as strong as originally thought, for all sources the linear correlation coefficient for the log of these parameters is 0.57, with a probability of this arising by chance of $<$2$\times$10$^{-7}$.
On this diagram we mark the location of \src\ and, as previously noted \citep[e.g.,][]{Halpern2007}, it lies in the region occupied by Be star systems. However, the length of the spin period is rather short compared to its orbital period as was also commented on by \citet{Markwardt2009}. We additionally mark in Figure \ref{fig:stars} the location of the other HMXBs that we compare with \src\ below.

\subsection{Comparison of Orbital Modulation with Other Systems}

\subsubsection{Be X-ray Binaries with Sharp Outbursts}

The sharpness of the, presumably orbital, outbursts in \src\ is exceptional in terms of phase duration. Most Be star systems show orbital outbursts which have durations of a considerable fraction of the orbit, typically \sqig0.2 - 0.3 P$_{orb}$ \citep{Reig2011} rather than the \sqig0.05 duration for \src.
Two other Be star systems that have been reported to have unusually sharp flares are A 0538-66 in the LMC \citep[e.g.,][]{Ducci2022} and AX J0049.4-7323
in the SMC \citep[e.g.,][]{Ducci2019}. \a0538 is an extreme system that has the shortest reported spin period for an HMXB (69 ms), a rather short orbital period of \sqig16.6 days with an eccentricity of \sqig0.7, and it has displayed super-Eddington outbursts \citep[e.g.,][]{Skinner1982,Raj2017}. Thus, similar to \src, A 0538-66 also lies below the spin/orbital period correlation trend for Be star HMXBs. In contrast, AX J0049.4-7323 has a 755\,s spin period which is not exceptionally short for its 394 day orbital period. The orbital eccentricity of AX J0049.4-7323 is unknown, but has been suggested to be high \citep[e.g.,][]{Cowley2003,Ducci2019} to account for the sharpness of the periodic optical outbursts.
\citet{Coe2004} proposed that the circumstellar disk around the Be star in AX J0049.4-7323 is disrupted during periastron passage, and noted a possible similarity to \a0538.

We note, however, that the duration of the flares in \src\ in terms of days, rather than phase, may not be that unusual. See, for example, \citet{Kuehnel2015} who argue that timescales are driven by the physics of accretion disk formation and dissipation around the neutron star. Thus, a system with a long orbital period will automatically have outbursts with short phase durations.

\subsubsection{Comparison with GS 1843-02 (2S 1845-024)}

\begin{figure}
\includegraphics[width=14cm,angle=0]{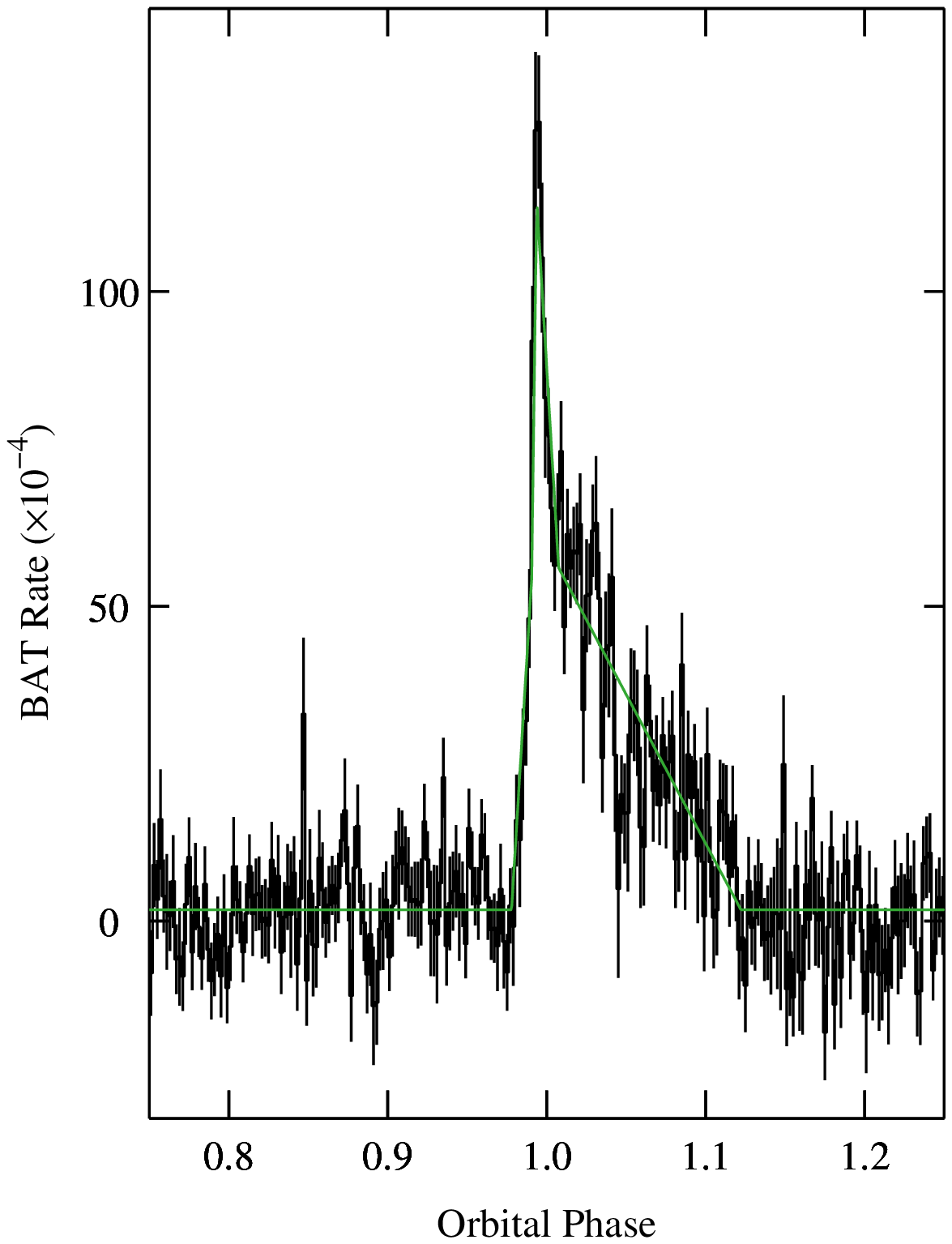}    
\caption{
Swift BAT light curve of GS 1843-02 (2S 1845-024) folded on its orbital period. The light curve was folded into 500 bins for one cycle, but only phases near periastron passage (0.75 to 1.25) are plotted. Phase zero corresponds to periastron passage. The green lines show the two triangular profiles used to characterize the outburst in this source that was fitted to the unfolded light curve.
}
\label{fig:gs1843_fold}
\end{figure}

GS 1843-02 (2S 1845-024) is a candidate Be star system that has a rather similar orbital period to \src\ 
at 242.18 days, although its spin period is much longer at \sqig95\,s, and its orbital eccentricity is very high at 0.869 \citep{Finger1999,Malacaria2020}. 
Long-term spin-up was reported by
\citet{Finger1999} from CGRO BATSE observations while \citet{Malacaria2020} using Fermi GBM data reported that GS 1842-02 had transitioned to showing long-term spin-down.
GS 1843-02 is also notable for having recurrent outbursts that have a very short phase duration. To compare with \src\ we examined the BAT light curve of GS 1843-02
that covers its most recent outburst state \citep{Malacaria2020} between MJD 53,416 and 58,214. This time period coincidentally corresponds to the quiescent Epoch II for \src, but since the two sources are separated on the sky by 8.\degrees74 no contamination between the light curves of the two sources is expected.
The folded profile of GS 1843-02 is shown in Figure \ref{fig:gs1843_fold}. In contrast to \src, the profile is clearly asymmetric with a much faster rise to maximum than decline back to the baseline rate. 
To characterize the outbursts in a simple way, we fit two periodic triangular outburst profiles
to the full non-folded light curve with the orbital period frozen at 242.1797 days. 
We find that the peak of the outburst occurs slightly before periastron passage ($\phi$ = 0)
with the sharper larger-amplitude component peaking 
1.8 $\pm$ 0.3 days ($\Delta\phi$ = 0.007 $\pm$ 0.001) earlier
while
the broader lower-amplitude component peaks at 1.5 $\pm$ 0.2 days ($\Delta\phi$ = 0.006) earlier.
This differs from the profile found by \citet{Finger1999} from BATSE observations which had a sharp peak approximately 0.5 days before periastron, with a broader symmetric profile peaking \sqig10 days after periastron. Hence GS 1843-02 experienced a change in outburst profile from the initial BATSE observations compared to the more recent activity state seen with BAT and the Fermi GBM. This is unlike \src\ where we find no large change in outburst profile between Epochs I and III. It is unclear whether the change in outburst profile for GS 1843-02 has any connection with the transition from long-term spin-up to spin-down in this source \citep{Finger1999,Malacaria2020}.

\subsubsection{Be Star Gamma-ray Binaries with Sharp Outbursts}

We also compare \src\ to a different type of binary, but which is likely evolutionarily connected to it.
High-Mass Gamma-ray Binaries (HMGBs) are thought to be progenitors of HMXBs, currently containing neutron stars that are rotating so rapidly that the pulsar mechanism inhibits accretion \citep{Meurs1989}.
We note that two Be HMGBs have high eccentricities, long orbital periods and only brief outbursts.
One is PSR B1259-63 which consists of a \sqig48 ms pulsar in orbit with a star of spectral type O9.5 \citep{Negueruela2011}.
It has an orbital period of 1237 days and an eccentricity of 0.87 \citep{Shannon2014}.
Away from periastron it appears as a radio pulsar, but near periastron it emits gamma-ray flares while the pulsations cease. The gamma-ray flares are generally ascribed to the interaction between the pulsar wind and the Be star disk \citep[e.g.,][]{Johnson2018,Coe2019}.
The other such HMGB is PSR J2032+4127 in which a 143 ms pulsar orbits a Be star with a period of \sqig 50 years and an eccentricity of 0.94 to 0.99 \citep{Ho2017}.
Once the pulsar in a high-eccentricity HMGB has spun down sufficiently to allow accretion to occur, it may thus become a high-eccentricity HMXB, which would be expected to have a longer lifetime.

\subsubsection{Magneto-centrifugal Inhibition of Accretion}

In order for accretion to occur onto a highly magnetized neutron star, it is necessary for the magnetospheric radius to be smaller than the corotation radius, otherwise material will be accelerated to super-Keplerian velocities which will inhibit accretion \citep[e.g.,][and references therein]{Stella1986,Corbet1996}.
The size of the magnetosphere depends on the accretion rate due to a balance between the ram pressure of the infalling material and magnetic pressure.
There can thus occur a large change in luminosity when the accretion rate moves above or below the critical rate at which the magnetospheric radius equals the corotation radius - a ``luminosity gap''.
\citet{Christodoulou2022} proposed that the detections at low flux levels seen with Chandra for \src\ outside the main outbursts (see Table \ref{table:flux_pulse_history})
fall at the lower end of this gap comparable to 4U 0115+63 and V 0332+53. In a simple model of the luminosity gap, the ratio of the minimum luminosity released by accretion onto the neutron star to the maximum energy released when accretion is halted at the magnetosphere is given by \citep{Corbet1996}:

\begin{equation}
\label{GapEquation}
\begin{split}
\Gamma & = L_{ns,min}/L_{m,max} \\
& = (G M_{ns})^{1/3} (P_{spin}/2 \pi)^{2/3}r_{ns}^{-1} \\
& \approx 150  \left(\frac{M_{ns}}{M_{\odot}}\right)\left(\frac{P_{spin}}{1s}\right)^{2/3}\left(\frac{r_{ns}}{10^6cm}\right)
\end{split}
\end{equation}

For \src\ this gives a ratio of \sqig660.
However, the actual transition between accretion onto the neutron star surface and accretion being blocked at the magnetosphere is likely to be more complex than in this simple model, see for example \citet{Postnov2017}.
In the outburst profiles from the RXTE PCA, BAT, and MAXI light curves we see approximately linear increases and decreases in count rates rather than sudden drops, however the size of the luminosity gap is too large for it to be detectable with these instruments.
Thus, any magnetic inhibition of accretion affecting the outburst profile seen with PCA, BAT and MAXI would be operating at a much lower level than a simple opening or closing of the accretion ``gate".

\section{Conclusion}

The periodic flares and long-term changes between active and inactive states seen in \src\ are typical of a Be star system. The location of the system in the P$_{spin}$/P$_{orb}$ diagram is also consistent with this classification. 
With a few exceptions, supergiant HMXBs have shorter orbital periods than Be systems and also do not
typically show periodic flaring. While GX 301-2 (4U 1223-62) is an exception with a longer orbital period of \sqig41.5 days and a significant eccentricity of \sqig0.47 with flares near periastron, it does not exhibit long-term changes in activity states \citep[e.g.,][and references therein]{Manikantan2024}. Thus, the evidence points toward \src\ containing a Be star. {\mybf In this case, if the primary is a main sequence B0 star, as discussed by \citet{Christodoulou2022}, the distance is \sqig4.5 kpc, and our values of $R_{\rm max}$ imply luminosities of \sqig4.6 and \sqig6.5 $\times10^{35}$ \ergss\ for Epochs I and III respectively which are typically for a Be X-ray binary \citep[e.g.][]{Reig2011}.
}

The periodic flares in \src\ are very short in phase duration and can be approximated with a rather symmetric triangular profile. Although, since the orbital period is relatively long, the outburst duration in time is not so short. By comparison with GS 1843-02 and \a0538, which are known to have high eccentricities, and AX J0049.4-7323 which is also suspected to have high eccentricity, this may also hint at a high system eccentricity for \src. Future pulse timing measurements of \src\ over the course of a complete outburst could provide a measurement of system eccentricity, which would help to establish its similarity to these sources. A determination of the spectral type of the primary in \src\ from near-infrared spectroscopy would aid in understanding the system, and such spectroscopy would show the angle of the Be star disk to our line-of-sight. Monitoring of near-IR spectral and photometric changes over the course of an orbit would reveal whether periodic sharp flares also occur in this waveband, and monitoring on longer-timescales would be expected to show changes in the Be envelope as it switches between active and inactive states.

\begin{acknowledgements}
We thank the referee for useful comments. CM acknowledges funding from the Italian Ministry of University and Research (MUR), PRIN 2020 (prot. 2020BRP57Z) ``Gravitational and Electromagnetic-wave Sources in the Universe with current and next generation detectors (GEMS)'' and the INAF Research Grant ``Uncovering the optical beat of the fastest magnetised neutron stars (FANS)''.
This research made use of PCA scan light curves provided by C. Markwardt, Swift/BAT transient
monitor results provided by the Swift/BAT team, and MAXI data provided by RIKEN, JAXA and the MAXI team.
The work was supported in part by NASA under award number 80GSFC21M0002.
Astrophysics research at the Naval Research Laboratory is supported by the NASA Astrophysics Explorer Program.
\end{acknowledgements}

\facilities{RXTE, Swift, MAXI}

\appendix
\section{Investigation of Modulation on Twice the Orbital Period}
\label{section:double_period}
\begin{figure}
\includegraphics[width=14cm,angle=0]{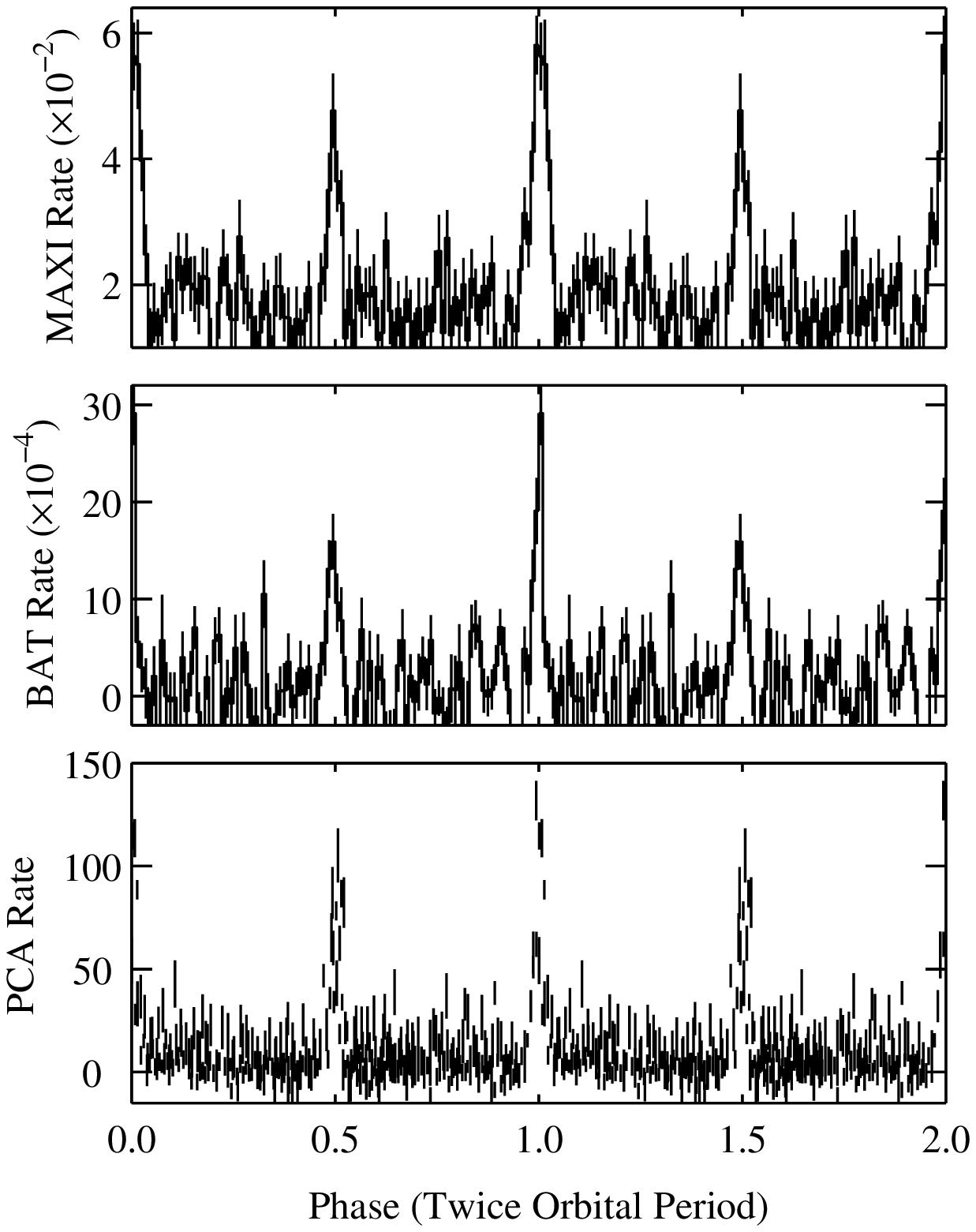}
\caption{
PCA (bottom), BAT (middle) and MAXI (top) light curves of \src\ folded on twice the assumed orbital period for Epochs I and III only.
}
\label{fig:double_fold}
\end{figure}

In the power spectra of the light curves from the active states (Figure \ref{fig:sel_power}) we noted
that the MAXI power spectrum shows a peak near twice the orbital period, with the BAT showing a peak close to this. We therefore investigated the light curves for the possibility of modulation on this period and folded them on twice the assumed orbital periods and these are shown in Figure \ref{fig:double_fold}. While all three instruments show an apparent difference between the peaks in the folded light curve at $\phi_{double}$ = 0.0 and $\phi_{double}$ = 0.5, it would be exceptional for an HMXB to show two very sharp flares per orbit. In addition, the profiles are only defined by a modest number of outbursts. We therefore suspect that the apparent odd/even difference is simply due to cycle-to-cycle variability and the orbital period is indeed \sqig244 days. Evidence for this also comes from the PCA light curve (Figure \ref{fig:pca_show_fits}) which shows that 
the apparently larger outbursts at $\phi_{double}$ = 0.0 compared to $\phi_{double}$ = 0.5 in Figure \ref{fig:double_fold} comes from a single outburst around MJD 53,460.

\end{document}